\documentclass[dvips]{article}

\usepackage{icrc2011}

\title{Detection of the Crab Pulsar with VERITAS above 100\,GeV}

\newcommand{\etal}{\MakeLowercase{\textit{et al. }}} 
\shorttitle{A.\ McCann \etal Detection of the Crab Pulsar with VERITAS above 100\,GeV}

\authors{A.\ McCann$^{1}$ for the VERITAS Collaboration$^{2}$}
\afiliations{$^1$Physics Department, McGill University, 3600 University Street
Montreal, QC H3A 2T8, Canada \\
$^2$see J. Holder et al. (these proceedings) or http://veritas.sao.arizona.edu/conferences/authors?icrc2011 }
\email{mccann@hep.physics.mcgill.ca}

\abstract{We discuss the recent detection of pulsed
gamma-ray emission from the Crab Pulsar above 100\,GeV with the
VERITAS array of atmospheric Cherenkov telescopes. Gamma-ray emission
at theses energies is not expected in present pulsar models. We find
that the photon spectrum of pulsed emission between 100\,MeV and
400\,GeV can be described by a broken power law, and that it is
statistically preferred over a power law with an exponential
cut-off. In the VERITAS energy range the spectrum can be described
with a simple power law with a spectral index of -3.8 and a flux
normalization at 150\,GeV that is equivalent to 1\% of the Crab Nebula
gamma-ray flux. The detection of pulsed emission above 100\,GeV and
the absence of an exponential cutoff rules out curvature radiation as
the primary gamma-ray-producing mechanism. The pulse profile exhibits
the characteristic two pulses of the Crab Pulsar at phases 0.0 and
0.4, albeit 2-3 times narrower than below 10\,GeV. The narrowing can
be interpreted as a tapered particle acceleration region in the
magnetosphere. Our findings require that the emission region of the
observed gamma rays be beyond 10 stellar radii from the neutron star.}
\keywords{ VHE, gamma-rays, pulsar, Crab, VERITAS }

\begin{document}
\maketitle

\section{Introduction}

One of the most powerful pulsars in gamma rays is the Crab Pulsar
\cite{12,13}, PSR J0534+220, which is the remnant of a historical
supernova that was observed in 1054 A.D. It is located at a distance
of 6500 light years, has a rotation period of $\approx$33 ms, a
spin-down power of $4.6\times10^{38}\,$erg s$^{-1}$ and a surface
magnetic field of $3.78\times10^{12}$\,G \cite{14}.

Within the corotating magnetosphere, charged particles are accelerated
to relativistic energies and emit non-thermal radiation from radio
waves through gamma rays. In general, gamma-ray pulsars exhibit a
break in the spectrum between a few hundred MeV and a few GeV. Mapping
the cut-off can help to constrain the geometry of the acceleration
region, the gamma-ray radiation mechanisms and the attenuation of
gamma-rays. Based on previous measurements and present theoretical
understanding that the dominant gamma-ray emission mechanism is
curvature radiation, it is widely believed that the shape of the
spectral break is best described by an exponential cut-off.

Although measurements of the Crab Pulsar spectrum are consistent with
a power law with exponential cut-off, flux measurements above 10\,GeV
are systematically above the best-fit model, suggesting that the
spectrum is indeed harder than a power law with exponential cut-off
\cite{13, 16}. However, the statistical uncertainty of the previous
data was insufficient to allow a definite conclusion about the
spectral shape. In this paper we summarise the recent detection of the
Crab Pulsar above 100\,GeV with VERITAS that rules out that the
spectrum above the break is described by an exponential cut-off. In
Section 2 we describe the observation and analysis. The results are
presented in Section 3 and we close the paper with a discussion in
Section 4.

\begin{figure*}[th]
  \centering \includegraphics[width=6.3in]{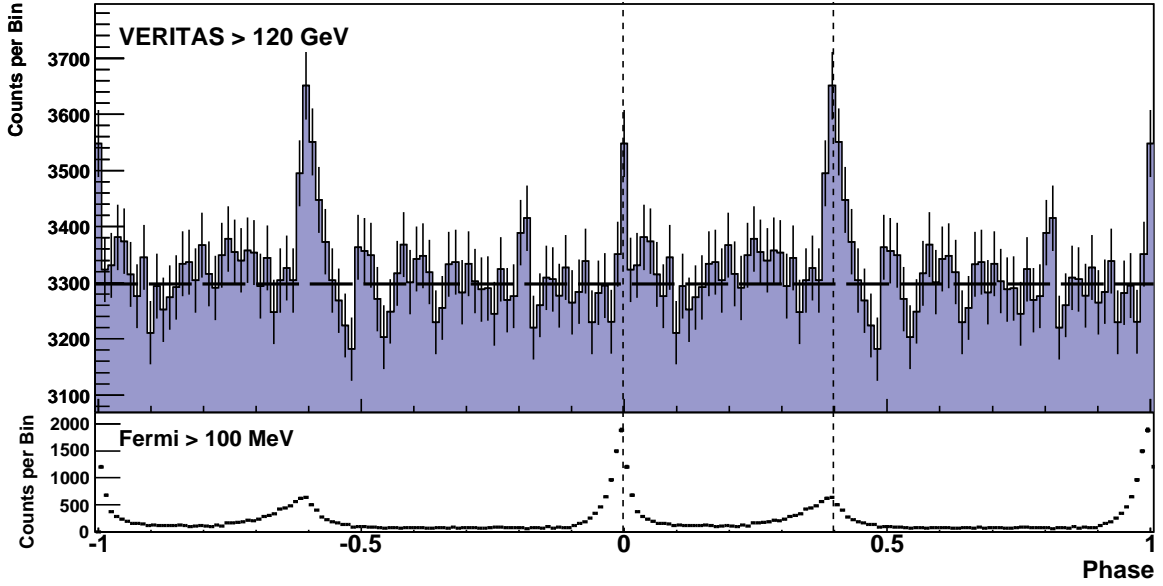}
  \caption{VERITAS pulse profile of the Crab Pulsar at
    $>120$\,GeV. The shaded histograms show the VERITAS data. The
    pulse profile is shown twice for clarity. The dashed horizontal
    line shows the background level estimated from data in the phase
    region between 0.43 and 0.94. The data above 100 MeV from the
    Fermi-LAT \cite{13} are shown beneath the VERITAS profile. The
    vertical dashed lines in the panels mark the best-fit peak
    positions of P1 and P2 in the VERITAS data.  }
  \label{profile}
 \end{figure*}

\section{Observation and analysis}

VERITAS, the Very Energetic Radiation Imaging Telescope Array System,
is an array of four, 12\,m diameter imaging atmospheric Cherenkov
telescopes located in southern Arizona, USA \cite{18}. After evidence
for pulsed emission was seen in 45 hours of data from the Crab Pulsar
that were recorded between 2007 and 2010, a deep 62-hour observation
was carried out on the Crab Pulsar between September 2010 and March
2011. The observations were made in “wobble” mode with a 0.5 degree
offset. After eliminating data taken under variable or poor sky
conditions or affected by technical problems, the total analysed data
set comprises 107 hours of observations (97 hours dead-time corrected)
carried out with all four telescopes. The data were taken with the
standard VERITAS trigger setting, and analysed with the standard
VERITAS analysis tools.

\begin{figure}[!t]
  \vspace{5mm}
  \centering
  \includegraphics[width=3.in]{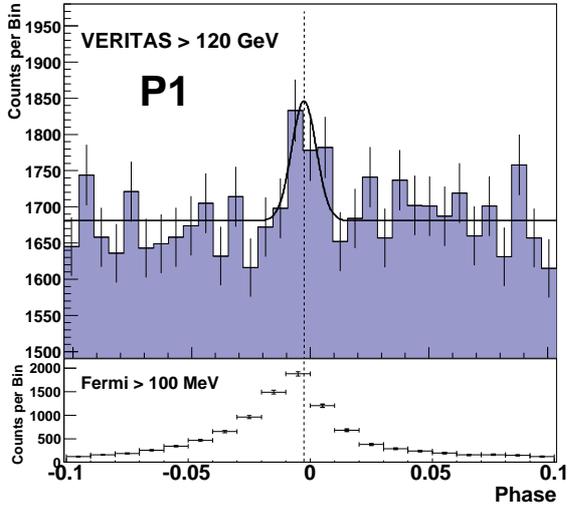}
  \caption{Enlarged view of the main pulse or P1 of the Crab Pulsar
    pulse profile. See text.}
  \label{zoomP1}
 \end{figure}

 \begin{figure}[!t]
  \vspace{5mm}
  \centering
  \includegraphics[width=3.in]{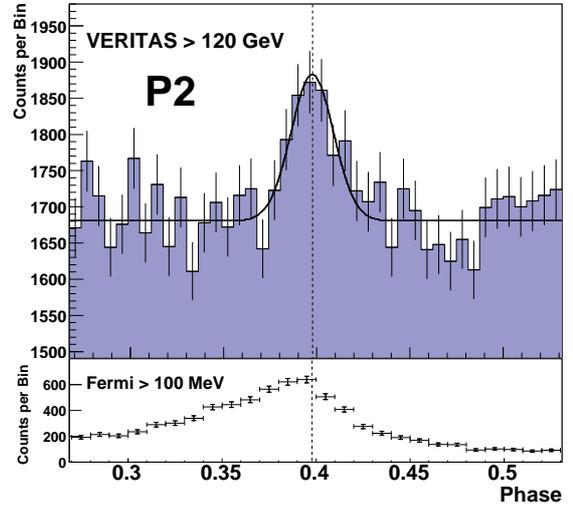}
  \caption{Enlarged view of the interpulse or P2 of the Crab Pulsar pulse profile. See text.}
  \label{zoomP2}
 \end{figure}

\subsection{Event Reconstruction}
In the analysis, the recorded atmospheric shower images are processed
with a standard moment analysis \cite{19} and the energy and arrival
direction of the primary particle are calculated \cite{20}. After
event reconstruction, a selection is performed to reject events caused
by charged cosmic rays. The selection criteria were optimised a priori
for highest sensitivity by assuming a simple power-law energy spectrum
for the Crab Pulsar with an index $\alpha = -4$ and a flux
normalisation of a few percent of the Crab Nebula flux at
100\,GeV. The analysis threshold is 120\,GeV.

For the pulsar analysis, the arrival times of the selected events are
transformed to the barycenter of the solar system.  After
barycentering, the phase of the Crab Pulsar is calculated for each
event using contemporaneous ephemerides of the Crab Pulsar that are
published monthly by the Jodrell Bank telescope \cite{21}.

The results were confirmed by a separate analysis of the data made
using an independent analysis package.

\section{Results}

\subsection{Pulse Profile}

\begin{figure*}[th]
  \centering \includegraphics[width=5in]{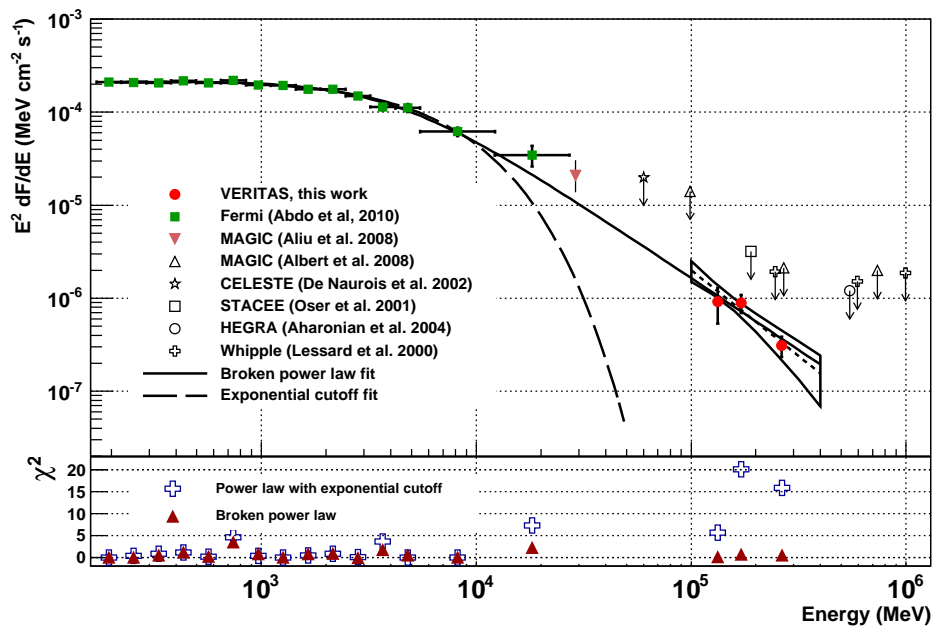}
  \caption{Spectral energy distribution (SED) of the Crab Pulsar in
    gamma rays. VERITAS flux measurements are shown by the solid red
    circles, Fermi-LAT data \cite{13} by green squares, and the MAGIC
    flux point \cite{16} by the solid triangle. The empty symbols are
    upper limits from CELESTE, MAGIC, and Whipple. The bow tie and the
    enclosed dotted line give the statistical uncertainties and the
    best-fit power-law spectrum for the VERITAS data using a
    forward-folding method. The result of a fit of the VERITAS and
    Fermi-LAT data with a broken power law is given by the solid line
    and the result of a fit with a power-law spectrum multiplied with
    an exponential cut-off is given by the dashed line. Below the SED
    we plot $\chi^2$ values to visualise the deviations of the
    best-fit parametrisation from the Fermi-LAT and VERITAS flux
    measurements.  }
  \label{sed}
 \end{figure*}

The phase-folded event distribution, hereafter pulse profile, of the
selected VERITAS events is shown in Figure \ref{profile}. The most
significant structures are two pulses with peak amplitudes at phase
0.0 and phase 0.4. These coincide with the locations of the main pulse
and interpulse, hereafter P1 and P2, which are the two main features
in the pulse profile of the Crab Pulsar throughout the electromagnetic
spectrum. In order to assess the significance of the pulsed emission,
we use the H-Test \cite{22}. The test result is 50, which translates
into a statistical significance of 6.0 standard deviations that pulsed
emission is present in the data.

The pulse profile has been characterised by an unbinned
maximum-likelihood fit; see the solid black line in Figures
\ref{zoomP1} and \ref{zoomP2}. In the fit, the pulses are modeled with
Gaussian functions, and the background is determined from the events
that fall between phases 0.5 and 0.9 in the pulse profile (referred to
as the off-pulse region). 

The positions of P1 and P2 in the VERITAS data are thus determined to
lie at the phase values $-0.0026\pm0.0028$ and $0.3978\pm0.0020$,
respectively and are shown by the vertical lines Figure 1. The full
widths at half maximum (FWHM) of the fitted pulses are
$0.0122\pm0.0035$ and $0.0267\pm0.0052$, respectively. The pulses are
narrower than those measured by Fermi-LAT at 100 MeV by a factor of
two to three. The energy-dependent narrowing of the pulses is a strong
probe of the nature of the magnetospheric particle acceleration region
and can be used to shed some light on its geometry, electric field,
and gamma-ray emission properties. If gamma rays observed at the same
phase are emitted by particles that propagate along the same magnetic
field line \cite{23}, then a possible explanation of the observed
narrowing is that the region where acceleration occurs tapers towards
the neutron star. However, detailed calculations are necessary to
explain fully the observed pulse profile.

\subsection{Spectral differences between P1 and P2}

Along with the observed differences in the pulse width, the amplitude
of P2 is larger than P1 in the profile measured with VERITAS, in
contrast to what is observed at lower gamma-ray energies where P1
dominates (see Figure \ref{profile}). It is known that the ratio of
the pulse amplitudes changes as a function of energy above 1 GeV
\cite{13} and becomes near unity for the pulse profile integrated
above 25\,GeV \cite{16}.

In order to quantify the relative intensity of the two peaks above
120\,GeV, we integrate the excess between phase $-0.013$ and 0.009 for
P1 and between 0.375 and 0.421 for P2. This is the $\pm$2 standard
deviation interval of each pulse as determined from the
maximum-likelihood fit. The significance of the excess in these
regions is 4.7 standard deviations for P1 and 7.9 standard deviations
for P2. The ratio of the excess events and thus the intensity ratio of
P2/P1 is $2.4\pm0.6$. If one assumes that the differential energy
spectra of P1 and P2 above 25\,GeV can each be described with a power
law, $dN/dE\propto E^{\alpha}$, and that the intensity ratio is
exactly unity at 25\,GeV \cite{16}, then the spectral index $\alpha$
of P1 must be smaller than the spectral index of P2 by
$\alpha_{\mbox{P2}} - \alpha_{\mbox{P1}} = 0.56 \pm 0.16$.

\subsection{Phase-averaged spectrum}

The gamma-ray spectrum above 100\,GeV was measured by combining the
signal regions around P1 and P2 defined above. This can be considered
a good approximation of the phase-averaged spectrum since no ``bridge
emission", which is observed at lower energies, is seen between P1 and
P2 in the VERITAS data. However, the existence of a flux component
that originates in the magnetosphere and is uniformly distributed in
phase cannot be excluded and would be indistinguishable from the
gamma-ray flux from the nebula. Figure \ref{sed} shows the VERITAS
phase-averaged spectrum together with measurements made with Fermi-LAT
and MAGIC. In the energy range between 100\,GeV and 400\,GeV measured
by VERITAS, the energy spectrum is well described by a power law
$dN/dE = A\times(E/150\,\mbox{GeV})^{\alpha}$, with $A = (4.2 \pm
0.6_{\mbox{stat}} +2.4_{\mbox{syst}} -1.4_{\mbox{syst}}) \times
10^{-11}$ TeV$^{-1}$ cm$^{-2}$ s$^{-1}$ and $\alpha = -3.8 \pm
0.5_{\mbox{stat}} \pm 0.2_{\mbox{syst}}$. The detection of pulsed
gamma-ray emission between 200\,GeV and 400\,GeV, the highest energy
flux point, is only possible if the emission region is at least 10
stellar radii from the star's surface \cite{24}.

\subsection{The spectral energy distribution between 100\,MeV and 300\,GeV}

Combining the VERITAS data with the Fermi-LAT data we can place a
stringent constraint on the shape of the spectral turnover. The
previously favoured spectral shape of the Crab Pulsar above 1\,GeV was
an exponential cut-off $dN/dE = A\times(E/E_0)^{\alpha}\exp(-E/E_C)$,
which is a good parametrisation of the Fermi-LAT \cite{13} and MAGIC
\cite{16} data. We note that the Fermi-LAT and MAGIC data can be
equally well parametrised by a broken power law but those data are not
sufficient to distinguish significantly between a broken power law and
an exponential cut-off. The VERITAS data, on the other hand, clearly
favour a broken power law as a parametrisation of the spectral
shape. The fit of the VERITAS and Fermi-LAT data with a broken power
law of the form $A\times(E/E_0)^{\alpha}/[1 + (E/E_0)^{\alpha-\beta}]$
results in a $\chi^2$ value of 13.5 for 15 degrees of freedom with the
fit parameters $A = (1.45 \pm 0.15_{\mbox{stat}}) \times 10^{-5}$
TeV$^{-1}$ cm$^{-2}$ s$^{-1}$, $E_0 = 4.0 \pm 0.5_{\mbox{stat}}\,$GeV,
$\alpha = -1.96 \pm 0.02_{\mbox{stat}}$ and $\beta = -3.52 \pm
0.04_{\mbox{stat}}$ (see solid black line in Figure 2). A
corresponding fit with a power law and an exponential cut-off yields a
$\chi^2$ value of 66.8 for 16 degrees of freedom. The fit probability
of $3.6 \times 10^{-8}$ derived from the $\chi^2$ value excludes the
exponential cut-off as a viable parametrisation of the Crab Pulsar
spectrum.

\section{Discussion}

The detection of pulsed gamma-ray emission above 100 GeV provides
strong constraints on the gamma-ray radiation mechanisms and the
location of the acceleration regions. For example, the shape of the
spectrum above the break can not be attributed to curvature radiation
because that would require an exponentially shaped cut-off. Assuming a
balance between acceleration gains and radiative losses by curvature
radiation, the break in the gamma-ray spectrum is expected to be at
$E_{br} = 24\,$GeV $\eta^{3/4} \sqrt{\xi}$, where $\eta$ is the
acceleration efficiency ($\eta < 1$) and $\xi$ is the radius of
curvature in units of the light-cylinder radius \cite{25}. Though
$\xi$ can be larger than one, only with an extremely large radius of
curvature would it be possible to produce gamma-ray emission above
100\,GeV with curvature radiation. It is, therefore, unlikely that
curvature radiation is the dominant production mechanism of the
observed gamma-ray emission above 100\,GeV. Two possible
interpretations are that either the entire gamma-ray production is
dominated by one emission mechanism different from curvature radiation
or that a second mechanism becomes dominant above the spectral break
energy.

\section{Acknowledgements}
This research is supported by grants from the US Department of Energy,
the US National Science Foundation, and the Smithsonian Institution,
by NSERC in Canada, by Science Foundation Ireland, and by STFC in the
UK. We acknowledge the excellent work of the technical support staff
at the FLWO and the collaborating institutions in the construction and
operation of the instrument.

\clearpage

\end{document}